# THIRD GENERATION RESIDUAL GAS IONIZATION PROFILE MONITORS AT FERMILAB*

J.R. Zagel[#], M. Alvarez, B. Fellenz, C. Jensen, C. Lundberg, E. McCrory, D. Slimmer, R. Thurman-Keup, D. Tinsley, FNAL, Batavia, IL 60510, USA

*Abstract*

The latest generation of IPM's installed in the Fermilab Main Injector and Recycler incorporate a 1 kG permanent magnet, a newly designed high-gain, rad-tolerant preamp, and a control grid to moderate the charge that is allowed to arrive on the anode pick-up strips. The control grid is intended to select a single Booster batch measurement per turn. Initially it is being used to allow for a faster turn-on of a single, high-intensity cycle in either machine. The expectation is that this will extend the Micro Channel Plate lifetime, which is the high-cost consumable in the measurement system. We discuss the new design and data acquired with this system.

## INTRODUCTION

Previous generations of residual gas ionization profile monitor's (IPM's) have been in operation, both at Fermilab and other laboratories, for many years [1-3]. The second-generation system applied a fixed permanent magnetic field, and polarity change to collect electrons to reduce space charge effects in the measurement [4]. Its field quality, while sufficient to improve the measurements, was not quite 1 kG, and required 4 feet of beam line to accommodate the magnetic unit. Two primary issues that arise have been the Micro Channel Plate (MCP) lifetime, and saturation effects of too much charge required from them. The third generation is designed to improve on these deficiencies.

## MEASUREMENT TIMING

The new system uses a timing module that combines the function of our Universal Clock Decoder (UCD), and the VME RF Timing module (VRFT). This module decodes the lab-wide timing 10 MHz clock (TCLK), the machine specific, beam synchronous clock (BSYNC), and a Machine Data (MDAT) Reference System, and outputs a clock trigger. Utilizing this board, we can generate a trigger on a specific machine state, and a particular machine clock cycle that then enables the beam synchronous clock pulses to gate our ADC's. The resolution of this trigger is one 53 MHz bunch. One sample per turn is generally acquired for a range of turns specified in the measurement specification. Multiple samples per turn is also a supported mode of operation.

## WHAT'S NEW

The new system incorporates the improvements added to the Fermilab Tevatron IPM [5] along with a few new features.

*New Magnet Design*

The new magnet is designed using $SmCo_5$ permanent magnets. This material was chosen to give the highest residual magnetic field (Br) with the lowest variation in temperature dependence, at a reasonable cost and availability. A standard brick is 1" by 2" by 1/2". The magnet was intended to provide 1 kG in the detector area, with the return flux half upstream and half downstream, and shimmed, such that the beam sees a total integrated field of zero. The magnet is 31.5"L x 14"D with a minimum aperture of 4.25" to allow for a 4" diameter beam pipe suitable to both horizontal and vertical planes in the Main Injector and Recycler. Magnet design, modeling, and measurement details have been presented in [6].

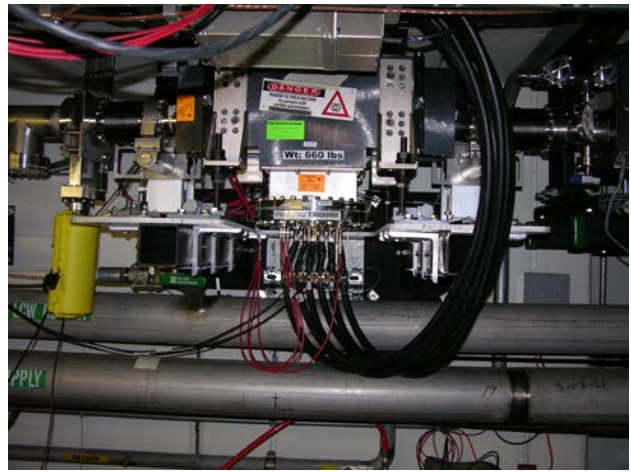

Figure 1: Recycler Vertical Installation

*Anode Strip Board*

The anode strip pick-up board for both Main Injector and Recycler is a ceramic circuit board with 120 copper strips spaced at 0.5mm. The copper occupies 80% of that pitch with 20% space between strips. All signals are connected via 20 pin, ceramic mounted, quick release connectors. A multilayer flexible circuit delivers these signals to 6 front flange feed through connectors. A "Flash Test" strip has been added to the bottom of the board to facilitate testing of the full analog signal path integrity. This strip is designed to be a 50Ω strip line that can be pulsed from outside of the tunnel.

### Control Grid

The third major improvement is the addition of a "Control Grid" that is placed on the beam side of the MCP's. The function of this control grid is to trap electrons and force them either upstream or downstream, outside of the detector area.

The grid material is electroformed nickel, 20 lines per inch with 0.00363" diameter wire, and open area of 86%. It is supported in a 316L stainless steel frame to cover the MCP aperture.

The control grid is kept at the full clearing field voltage until the batch of interest is passing through the detector. It is then pulsed to ground to allow the electrons to pass, un-deflected, on to the MCP. The capacitance of this grid is such that. With a 600 Ohm source impedance, the rise time of is 100 nsec. The external cable adds approximately 100 pF per meter.

Our batch structure is a string of 84 bunches, which represents a single full Booster turn. Up to 6 full Booster batches can be populated in Main Injector, leaving one equivalent 84 bunch length gap, for extraction kicker rise time. These 84 bunches constitute a 1.6 µsec batch. Ideally we plan to enable the grid slightly before the first bunch in the batch of interest arrives and shut it down after the last bunch passes.

### Pre-Amplifier

The fourth major improvement is a redesigned preamplifier that lives within 2 meters of the detector. The preamp input will see all of the charge induced onto the anode strips and amplifies it to a reasonable voltage for transmission outside the tunnel. It has radiation tolerant parts, and is mounted away from the plane of the beam to further reduce particle bombardment.

The amplifier was designed with a 100 Ohm input impedance, and a bandwidth of 1.2MHz providing sufficient rise time for the 1.6 µsec batch length, but also a high frequency signal rejection at 53 MHz. An OPA846, was chosen for the two stages for its low input voltage noise, high bandwidth, dc accuracy and output drive.

The amplifiers have an overall voltage gain of 5000 into 50 ohms and an input impedance of 100 Ohms that together form a trans-impedance of 500 kOhm. The analog front end is set to +/- 1.5 Volts full scale. At the input of the amplifier is a 1.2 MHz low pass filter to reduce the beam's 53 MHz RF signal that couples onto the strips, this was measured to be less than -80dB. The low bandwidth of the amplifier in effect integrates the charge of each batch (up to 84 populated bunches).

Twenty channels are placed on a single board, in a single wide NIM module with six of these modules in a NIM crate to make a total of 120 channels per IPM system. Coupling between channels has been measured to be less than -56dB. The crate was placed in the tunnel close to the IPM detector to reduce noise on the signals.

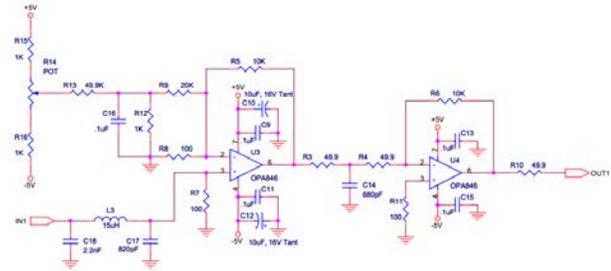

Figure 2: IPM Pre-Amplifier 2014 design

### Analog to Digital Converter.

A new series of VME Analog to Digital Converter's (ADC) has been designed by the Fermilab Accelerator Instrumentation Department. The boards produced for the IPM system are populated with 16 channels of 14 bit 80MHz parts. Also on board is an Field Programmable Gate Array (FPGA) that allows us great flexibility in our data acquisition. We sample multiple times during a gate that is centered on the batch of interest, and then average those samples to obtain a single recorded value for that batch for that particular turn. The sample depth is limited only by the amount of memory installed on the board, and the processing time desired for a result.

In addition, the FPGA also has an MDAT frame decoder that supplies our system with the intensity of the beam in the machine and the energy of the machine at the time that the analog data is sampled. A total of 96 Channels are digitized to create the beam profiles.

### LabVIEW™ Front End Software Enhancements

We have added new features to the third generation IPM program to enhance utility and usability.

The 2D display now has XY cursors enabled. When the Y cursor is moved, the new sample number is captured and used to initiate a single turn analysis. The X cursor destination (channel) is displayed below the X axis scale as an aid for identifying individual channels of interest.

There is now a running total stored and displayed that represents the amount of time high voltage has been applied to the micro channel plate. This number, in conjunction with the S/N (dB) is provided to help analyze micro channel plate lifetime.

A MCP accumulated plate charge correction button is available for the 2D display. This correction assumes there is no significant change in amplitude, sigma, or position from the reference sample (default sample 500). The correction is for visualization purposes only and is not used in the single or multi-sample analysis.

Channel data values above or below predetermined limits are now automatically eliminated from display in the Single Turn analysis plot. Limits are established using the specifications of the ADC and the expected normal data range. Channels outside this range are still represented in the 2D raw data display.

Position and sigma FFT data is now reported to ACNET devices in addition to the front panel display.

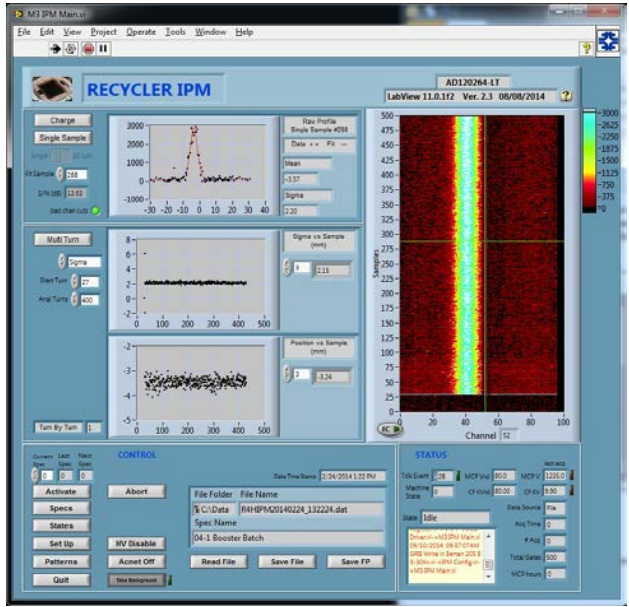

Figure 3: LabView™ User interface.

*Movable Detector*

The detector vacuum vessel is mounted on a pair of linear slides at each end that allow us to change the area of the MCP that is exposed to beam. Initially we can change the alignment of the detector to the beam by skewing the unit a few degrees from end to end. We take data to verify the alignment that yields the minimum sigma measured during stable beam conditions. This also facilitates moving the unit to a different spot on the MCP after every few weeks of operations.

There is a total travel of 3.8 cm centered on the proton beam. However, we can position the vacuum vessel much farther into the magnet to provide extra clearance when we need to remove the detector for maintenance.

Moving the vacuum vessel without changing the position of the magnet means that the beam is exposed to the same magnetic field.

*Maintenance Enhancement*

A major step forward is the addition of isolation valves immediately adjacent to the instrument both upstream and downstream. The detector can be isolated from the rest of the beam pipe for removal and replacement of the MCP. This minimizes the time for vacuum recovery after maintenance. A cross with a pump out port and an ion pump, is installed inside the isolation valves. To vent the instrument, we let it up to dry nitrogen before opening it to atmosphere.

## MEASUREMENT CAPABILITY

A total of 96 Channels are digitized to create the beam profiles. Depending on the number of cables coming up from the tunnel, some anode strips are skipped at the tails to provide for a good fit.

These instruments are generally used for injection matching. A detailed analysis is performed on the first 500 turns after injection to look for either sigma oscillations, indicating a betatron oscillation, or position oscillations, indicating beam was injected off orbit.

## OBSERVED PERFORMANCE

The control grid, at full clearing field voltage, is effective at diverting over 90% of the electrons formed by ionization from bunches other than the desired measurement. We currently have the ability to switch the control grid from our service buildings, allowing us to start a measurement soon after decoding the appropriate clock events. Turn on time is about 7.25 µsec. Revolution time in both Main Injector and Recycler is on the order of 11.1 µsec.

High voltage switches are available in the commercial market that will provide fast rise times at 10 kV for the low capacitance load of this grid. The peak current is limited to 12 Amps by a series resistance. These switches are built using MosFET technology, but are not radiation tolerant. Switching the control grid for the batch measurement is not yet possible as switching time is dominated by the capacitance of the 100-150 foot cables from service building to tunnel. We are looking for a place in an alcove where we could potentially build a somewhat shielded enclosure to house the switches.

## CONTROL GRID SIMULATION

The IPM simulation is a MATLAB program that tracks the ionization particles from their point of production to some arbitrary stopping point. It utilizes the time-dependent electric and magnetic fields of the proton bunches as they pass through the IPM region. It also includes the calculated magnetic field of the IPM magnet from OPERA, and the 2-D electric field of the IPM as calculated via POISSON. Ionization particles with energy $E$ are generated from a $1/E^2$ distribution. Figure 4 shows the drift of the ionization electrons when the IPM is gated on. This is the normal behaviour.

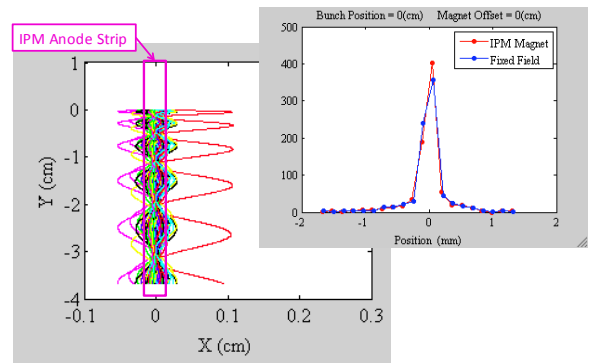

Figure 4: Plot of electron drift lines. All electrons start from the same point and take about 1.7 ns to drift to the bottom. The inset plot shows the distribution of particle positions at the bottom. Most of them are contained within a single anode strip.

When the grid is gated off, the field lines are as shown in Figure 5.

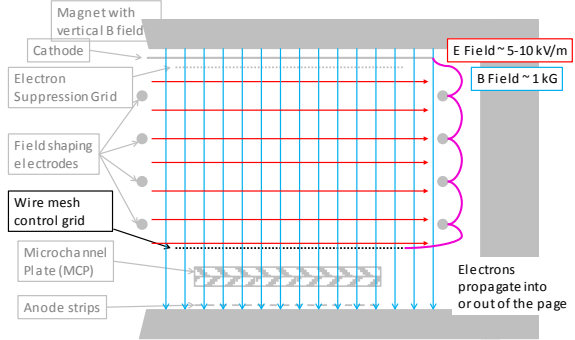

Figure 5: Field lines when the control grid is gated off. The crossed electric and magnetic fields produces motion in the perpendicular direction. In this case that is along the proton beam line.

The simulation results of the drift lines for the gated-off case are shown in Figure 6. The electron drift velocity along the proton beam line in the simulation is ~8 cm/μs compared to ~10 cm/μs estimated analytically assuming a uniform electric and magnetic field and no protons. This velocity implies that the ionized electrons will drift out of the region of the MCP in ~1 μs.

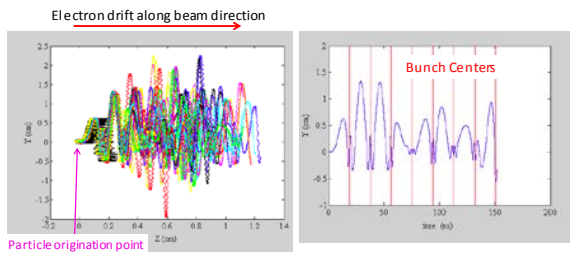

Figure 6: Drift trajectories of electrons with the control grid gated off. The plot on the right shows the trajectory of a single electron where one can see the constraining of the trajectory that occurs when a proton bunch passes by.

The drift trajectories of the ions are shown in Figure 7. Ions that encounter the grid are not directly a problem, however if they generate secondary electrons on the far side of the control grid, those would then accelerate toward the MCP and could partially negate the benefits.

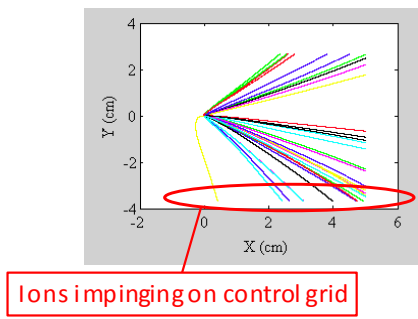

Figure 7: Ion trajectories with control grid gated off. Since they are heavy, they are relatively uninfluenced by the electric and magnetic fields. The elapsed time to reach the control grid was ~1.5 μs.

## CONCLUSION

The improvements detailed all contribute to the goal of making a better measurement while at the same time extending micro channel plate lifetime. We still have work to do to design a cost effective way of switching high voltages, at 100 kHz, in the Fermilab tunnels. This work continues.

## ACKNOWLEDGMENT

The authors would like to acknowledge all the work of the Fermilab Technical Division for their efforts in designing, building, and testing our new magnet for this project. In addition, we could not have gotten this far without the collaborations with other individuals at various laboratories as cited in the reference section below. Lastly, we acknowledge the efforts of many technicians who spent many hours in the assembly, testing, vacuum certification, and installation of the instruments.